# Photonuclear interactions at very high energies and vector meson dominance

E V Bugaev[1] and B V Mangazeev[2]
[1]Institute for Nuclear Research of the Russian Academy of Sciences,
7a, 60th October Anniversary prospect, Moscow 117312, Russia
[2]Irkutsk State University, 1, Karl Marx Street, Irkutsk 664003, Russia

E-mail: bugaev@pcbai10.inr.ruhep.ru

**Abstract.** We show that nucleon electromagnetic structure functions of deep inelastic scattering in Regge-Gribov limit (fixed $Q$-squared, asymptotically large $1/x$ and $s$) can be well described in the two-component (soft + hard) approach. In the concrete model elaborated by authors, the soft part of the virtual photon-nucleon scattering is given by the vector meson dominance, with taking into account the radial excitations of the rho-meson and nondiagonal transitions in meson-nucleon interactions. The hard part is calculated by using the dipole factorization, i.e., the process is considered as the dissociation of the photon into a $q\bar{q}$-pair (the "color dipole") and the subsequent interaction of this dipole with the nucleon. The dipole cross section has a Regge-type $s$-dependence and vanishes in the limit of large transverse sizes of the dipole. We give the brief description of the model and present results of the detailed comparison of model predictions with experimental data for electromagnetic structure functions of the nucleon.

## 1. Introduction

During last 10-15 years deep inelastic scattering (DIS) at small $x$ and not very small $Q^2$ ($Q^2>$1-2 GeV$^2$) has been modeled, mostly, by perturbative QCD (PQCD).

Nonperturbative theory, by definition, works, partly at least, in strong coupling regime and takes into account confinement effects. These effects are especially important at small and moderate $Q^2$, but are often ignored or incorporated purely phenomenologically. It is so, in particular, in QCD-inspired theories of a Pomeron where the description is based on the two-gluon exchange diagrams.

Clearly, nonperturbative effects in DIS must be taken into account because the $q\bar{q}$- pairs (produced, at first stage, by the virtual photon) with transverse sizes $r_\perp >$ 2 GeV$^{-1}$ are strongly overlapped with hadronic states (mostly, with vector mesons). It appears, however, that corresponding nonperturbative effects are partly masked by effects of gluon saturation (the latter effects are important if $Q^2 < Q_s^2$, where $Q_s$ is the saturation scale, and $Q_s^2(x) \sim$1.5-2.5 GeV$^2$ for $x \sim 10^{-5}$-$10^{-6}$). But, in general, this masking is not complete because kinematical region where saturation is important does not always coincide with region where confinement is essential.

It was stressed recently that, in spite of fact that modern PQCD approaches based, e.g., on Color Glass Condensate theory, are rather successful in describing the DIS data, the contribution of the nonperturbative physics to the $F_2$ structure function at small $Q^2$ may be non-negligible. It appears [1], in particular, that strong coupling based methods (like AdS/CFT [2]) are able to describe $F_2$ data at small $Q^2$ rather well.

The vector meson dominance (VMD) is the truly nonperturbative approach because it operates with hadrons and, consequently, the confinement effects are automatically built in. The most general formulation [3] of the VMD concept (referred to as the "Generalized VMD") uses the double mass dispersion relation for the forward Compton scattering amplitude $A_{\gamma p}$:

$$\frac{1}{s}\operatorname{Im} A_{\gamma p} = \int \frac{dM^2}{M^2+Q^2} \int \frac{dM'^2}{M'^2+Q^2} \rho(s,M^2,M'^2) \frac{1}{s} \operatorname{Im} A_{Vp \to V'p}. \qquad (1)$$

Here, $M$ and $M'$ are the invariant masses of the incoming and outgoing vector mesons, $\rho(s,M^2,M'^2)$ is the spectral function. The field-theoretical basis of this formula goes back to Sakurai's idea of treating the ρ-meson as gauge boson and to hidden-gauge theories, i.e., theories, in which the vector meson is a gauge boson of the hidden local symmetry (HLS).

In last several years a lot of papers appeared, in which holographic models of QCD and, moreover, holographic models of hadrons are elaborated. These models are extra-dimensional, essentially nonperturbative (in particular, color confinement is built in), based on AdS/QCD correspondence. Vector meson dominance and hidden local symmetry are natural consequences of these models. Important (for us) predictions of holographic QCD are as follows:

1. There are towers of vector resonances with infinite numbers of particles: ρ, ρ', ρ''... .
2. The mass spectrum of vector mesons depends on the geometry: $M^2 \sim n^2$ in "hard-wall" models or $M^2 \sim n$ in "soft-wall" models (see [4] for references).
3. There is the current-field identity: $J_V^\mu(x) = \sum_{n=1}^{\infty} f_V^n V_{(n)}^\mu(x)$.
4. The pion (as well as the nucleon) electromagnetic formfactor is completely meson dominated,

$$F_\pi(Q^2) = \sum_n \frac{f_V^{(n)} g_{n\pi\pi}}{M_n^2 - Q^2}.$$

The last two points show that one can, in some sense, say about the "return of vector dominance": the "old" vector dominance with the lowest $V_{(1)} = \rho$ is replaced everywhere by a "new", extended, vector dominance with an infinite tower of vector mesons.

## 2. Outline of the model

It is well known that the consistency of the spectral representation (1) with the approximate Bjorken scaling requires, in the approaches based on VMD, rather unnatural (from point of view of hadron physics) strong cancellations between amplitudes of the diagonal and nondiagonal, Vp→V'p, transitions. It had been shown in our previous works [4-7] that such cancellations are not effective. More exactly, there is no motivation for essential cancellations if it is assumed that the vector mesons of VMD models are similar to vector mesons with known properties. On the other hand, if it is assumed that these states are $q\bar{q}$-systems with definite mass, $M_{q\bar{q}}$, rather than vector mesons, the essential cancellations between diagonal and off-diagonal transitions become possible, due to a general feature of quantum field theory that fermion and antifermion couple with opposite sign (the well-known example is the case of two-gluon exchange). Therefore, in some modern versions of VMD there are no vector mesons at all, only $q\bar{q}$-pairs, i.e., these models are not hadronic.

As a way out of this situation we proposed the two-component model of DIS at small $x$ (the idea had been suggested in [5]): interactions of the $q\bar{q}$- pair (produced by the virtual photon) with the target nucleon can be described by PQCD if the transverse size $r_\perp$ of the pair is small and by VMD if $r_\perp$ is of the order of typical hadronic size. Correspondingly, in this model the structure functions of DIS have two components: the "soft" component described by VMD and the "hard" one described by PQCD. Naturally, VMD is used in its "aligned jet" version [8], i.e., the configurations are selected, in which the $q$ and $\bar{q}$, produced by virtual photon, are aligned along the beam direction and, as a consequence, the transverse distance between $q$ and $\bar{q}$, on arrival at the target nucleon, is of the order of hadronic size. Amplitudes of nondiagonal transitions (Vp-V'p) are calculated in the two-gluon exchange approximation, and the vector mesons are treated as bound states of quark and antiquark. The wave functions of these bound states are obtained from the solution of Bethe-Salpeter equation with an effective input kernel having the long-range confining part.

For a description of the perturbative (hard) component we use the colour dipole model with the dipole cross section having a Regge type $s$-dependence. For the latter we use the parameterization suggested by Forshaw, Kerley and Shaw [9, 10] (omitting the soft part of it). For dipoles with small transverse size the dipole cross section has a behavior $\sim r^2 (r^2 s)^{v_H}$ as $r$ goes to zero. We use the FKS formula for calculations of nucleon structure functions without any modification except the region of extremely small $x$ ($x<10^{-5}$). Namely, we assumed that $v_H$ slightly increases with a decrease of $x$ ($v_H$=3.27, $v_H$=4, $v_H$=5 at, correspondingly, $x$=$10^{-5}$, $x$=$10^{-7}$, $x$=$10^{-9}$).

Note, at the end of this section, that similar two-component models of DIS (using the VMD approach in its simplest form, without radial excitations and nondiagonal transitions) had been developed in [11, 12].

## 3. Main results of the calculations

The main results of our calculations are presented in figures 1-4. Figure 1 shows an energy dependence of the photoabsorption cross section for the real photon. We assumed that the soft part of $\sigma_{\gamma p}(s)$ can be parameterized by the Regge type formula

$$\sigma_{\gamma p}(s) = 114 \left( \frac{1.15}{\sqrt{s}} + \left( \frac{s}{1700} \right)^{0.06} \right) \quad (2)$$

($s$ in GeV$^2$, $\sigma_{\gamma p}(s)$ in microbarns). At small $s$ the soft part of $\sigma_{\gamma p}(s)$ completely dominates. For the hard component of $\sigma_{\gamma p}(s)$ the $Q^2$=0 limit of FKS formula is used, with the assumption that square of the mass of the quark is equal to 0.08 GeV$^2$.

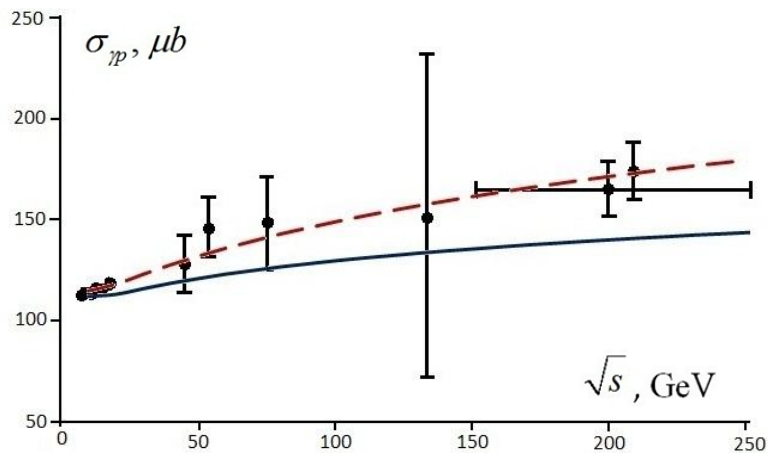

**Figure 1.** The total cross section of photoabsorption for the real photon (the dashed line). The solid line is the soft contribution (see [4] for references to the experimental data points).

Figures 2a, 2b show the $Q^2$-dependences of the structure function $F_2$, for fixed values of $x$. It can be seen from these figures that the model describes the data satisfactorily in the region $Q^2<10$ GeV$^2$, $x<0.05$. At larger $Q^2$ the hard component dominates.

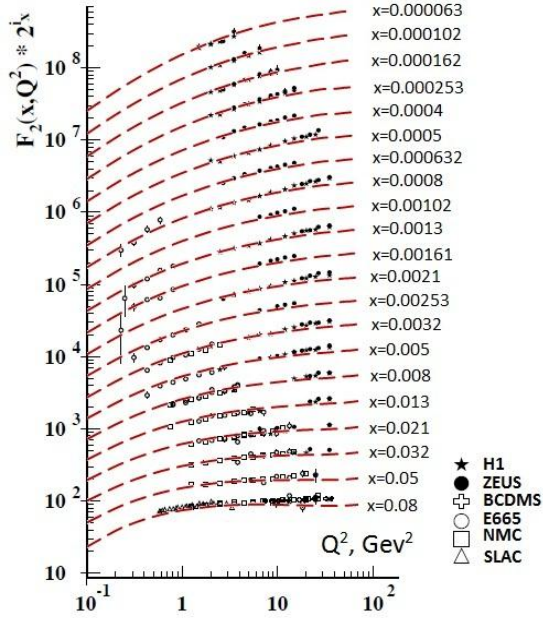
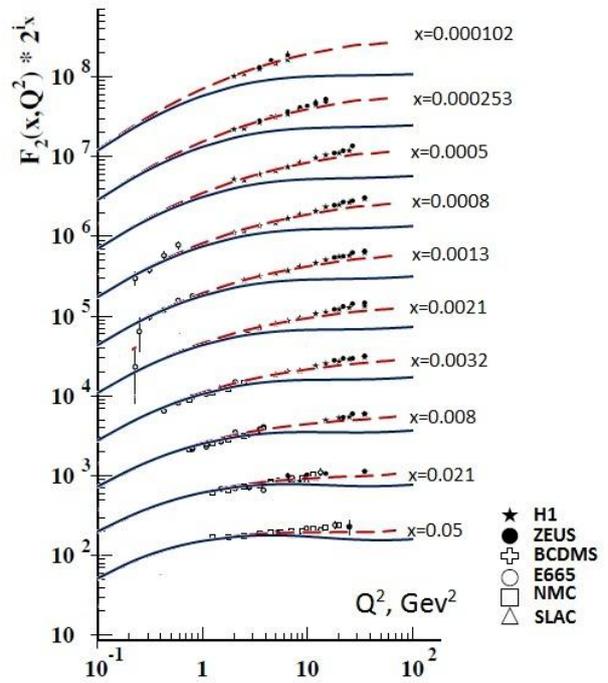

**Figure 2a.** The $Q^2$ dependence of the structure function $F_2$ for different values of $x$. Data of each bin of fixed $x$ has been multiplied by $2^i$, where $i$ is the number of the bin, ranging from $i=8$ ($x=0.08$) to $i=28$ ($x=0.000063$). The experimental points are taken from [13].

**Figure 2b.** The same as figure 2a except that the data of $x$-bins are shown for the bins with odd numbers only, ranging from $i=9$ ($x=0.05$) to $i=27$ ($x=0.000102$). The solid lines are the soft contributions.

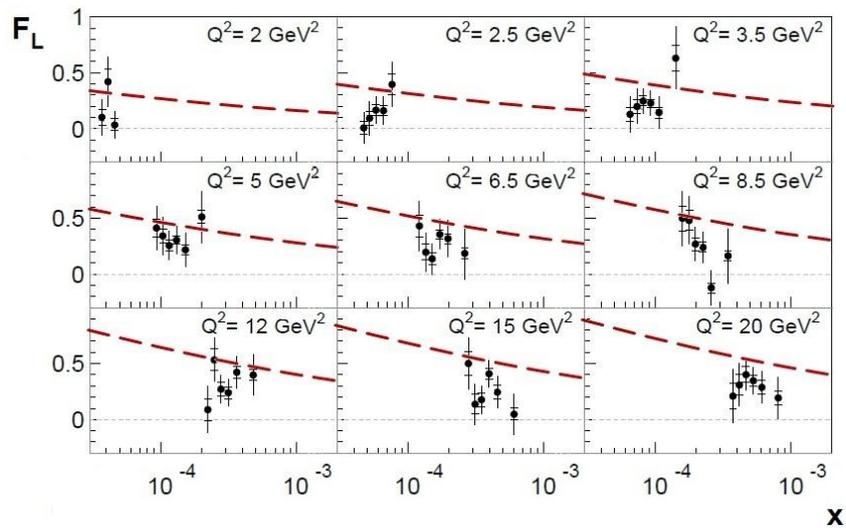

**Figure 3.** The $x$ dependence of the structure function $F_L$ for different values of $Q^2$. The experimental points are taken from [14] (H1 Collaboration).

In figure 3 the comparison of our predictions with data on longitudinal structure function $F_L$ is shown. The soft part of $F_L$ strongly depends on the assumptions on the ratio of longitudinal to transverse vector meson absorption cross sections, $\xi = \sigma_L / \sigma_T$, which is the parameter of VMD model (due to an absence of experimental data at large energies). We assumed that $\xi$ goes to 1 at $s$ goes to infinity (according with the $s$-channel helicity conservation) and parameterized the $\xi(s)$ dependence by the formula $\xi(s) = 1 - 0.9\exp\left(-(s/10000)^{1/4}\right)$, where $s$ in GeV$^2$. The comparison with scarce data in figure 3 shows that the model overestimates $F_L$, especially at large $Q^2$.

Figure 4 shows the $x$-dependences of $F_2$ for fixed values of $Q^2$ in the region of very small $x$, and rather small $Q^2$ (the region, which is important for experiments with cosmic rays). One can see that $F_2$ slowly increases with a decrease of $x$, while a relative contribution of the soft component decreases (although even at $x\sim10^{-9}$ and $Q^2\sim1$ GeV$^2$ this contribution is not too small, ~40 %).

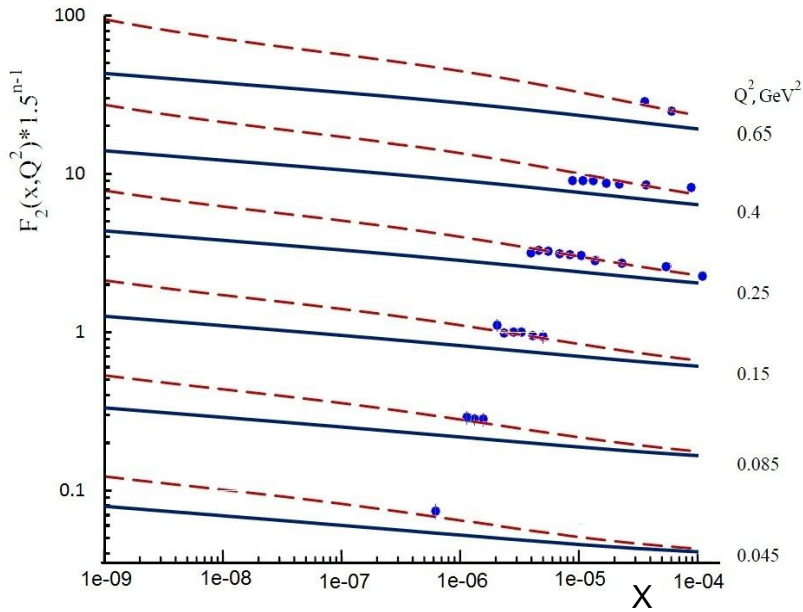

**Figure 4.** The $x$ dependence of the structure function $F_2$ for small values of $Q^2$ in the region of very small $x$. The experimental points are taken from [13]. The data and lines are scaled by powers of 1.5 from bottom to top ($n$ = 1, 3, 5 …). The solid lines are the soft contributions.

## 4. Conclusions
We demonstrated that the two-component approach to the theoretical description of diffractive DIS, in which the nonperturbative part is given by the VMD in its modern form (suggested by the AdS/CFT correspondence) is successful in description of experimental data at small $x$ ($x<0.08$) and $Q^2\leq10$ GeV$^2$. This may be enough for using in cosmic rays experiments if one uses muons from the steeply falling cosmic ray muon spectrum.

At $Q^2>10$ GeV$^2$ the perturbative part of $F_2$ becomes dominant (for $x\sim10^{-4}$) in our calculations. Note, once more, that we use for the perturbative part the hard Pomeron piece of colour dipole cross section from the works [9,10] (without any modification, in the region $x>10^{-6}$, where the most of data exists). We showed also that in the region of extremely small $x$, smaller than $10^{-6}$, and at $Q^2$ around 1 GeV$^2$ (this region is especially important for applications in cosmic ray physics) the satisfactory description of structure functions of DIS can be obtained within a framework of the two-component (PQCD + VMD) approach (and, moreover, both components are equally essential).